\documentclass[aps,prd,groupedaddress]{revtex4}

\usepackage{graphicx}
\usepackage{epstopdf}

\newcommand\fdg{\mbox{$.\!\!^\circ$}}

\begin{document}

\title{Phase-space distribution of unbound dark matter near the Sun}

\author{Moqbil S. Alenazi}
\email{alenazi@physics.utah.edu}
\author{Paolo Gondolo}
\email{paolo@physics.utah.edu}
\affiliation{Department of Physics, University of Utah, 115 S 1400 E Rm 201, Salt Lake City, Utah 84112-0830, USA}

\begin{abstract}
We resolve discrepancies in previous analyses of the flow of collisionless dark matter particles in the Sun's gravitational field. We determine the phase-space distribution of the flow both numerically, tracing particle trajectories back in time, and analytically, providing a simple correct relation between the velocity of particles at infinity and at the Earth. We use our results to produce sky maps of the distribution of arrival directions of dark matter particles on Earth at various times of the year. We assume various Maxwellian velocity distributions at infinity describing the standard dark halo and streams of dark matter. We illustrate the formation of a ring, analogous to the Einstein ring, when the Earth is directly downstream of the Sun.
\end{abstract}

\maketitle

\section{Introduction}
\label{introduction}

The motion of galaxies in the Universe cannot be explained by the gravitational pull of visible matter. Invisible dark matter (DM) is introduced to account for galactic rotation velocities and the gravitational interaction between galaxies. The spinning motion of the spiral disk of our Milky Way Galaxy is too fast to be explained merely by the gravity of its visible stars and gases, so it is assumed that the disk is surrounded by a much larger halo that contains a few stars but mostly unseen DM. A galaxy like our Milky Way is about $10 \%$ visible matter (stars and gas) and $90 \%$ unseen DM. 

The nature of DM is one of the unsolved problems of astrophysics and cosmology. Numerous theories have been proposed regarding the nature of the DM, from brown dwarf stars to MACHOs (Massive Compact Halo Objects) and theoretical subatomic particles such as WIMPs (Weakly Interacting Massive Particles) or axions. More than 20 groups of physicists worldwide have been building devices to detect DM particles. 

Several of these devices aim to detect the elastic scattering of WIMPs off a laboratory target. Some of them also attempt to measure the direction from which the WIMPs reach the detector. These directional searches will most probably be the only observational way to explore the astrophysical properties of WIMPs near the Solar System. Current theoretical expectations for the local WIMP distribution range from a virialized Maxwellian velocity function (isothermal model) to a more or less complex network of streams of WIMPs. These streams arise either from the tidal disruption of satellite galaxies, like the Sagittarius dwarf, or from the process of gravitational collapse during the formation of the Milky Way Galaxy.

For the planning and future interpretation of WIMP searches, it is important to identify and analyze the density and velocity distribution of DM particles in the Solar System. Damour and Krauss \cite{DK}, Gould \cite{Gol}, Gould and Alam \cite{GA}, and Lundberg and Edsjo \cite{LE} analyzed the transfer of DM particles from unbound to bound orbits caused by gravitational collisions with planets and by scattering with nuclei inside the planets or the Sun. Previously, Danby and Camm \cite{DC}, Danby and Bray \cite{DB}, Griest \cite{G}, and Sikivie and Wick \cite{SW} studied the distribution of particles on unbound orbits, but their conclusions disagree with each other. All these papers assumed an isotropic Maxwellian distribution of velocities for the particles at infinity. 

To resolve the discrepancies in the latter studies just mentioned, and to prepare for an analysis of velocity distributions more general than an isotropic Maxwellian, this paper is devoted to clarify the effect of the gravitational field of the Sun on the distribution of unbound orbits in our Solar System. We neglect the effects of the planets on the DM particle distribution.

\begin{figure}[t]
\centering
\includegraphics[width=5.5in]{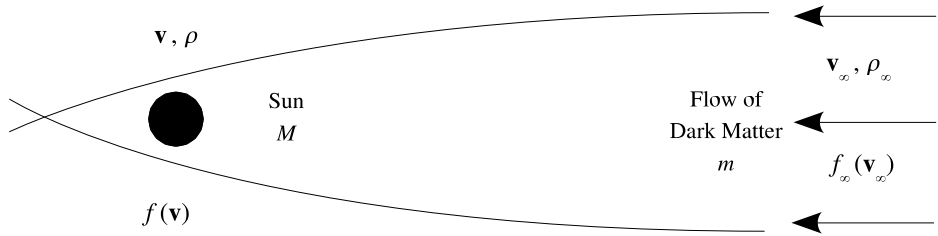}
\caption {Illustration of a flow of dark matter particles through the gravitational field of the Sun.}
\label{first}
\end{figure}

As an illustration of the problem we address, Fig.~\ref{first} depicts a flow of DM particles coming from infinity with a mean velocity ${\bf V}$ relative to the Sun and a velocity distribution function $f({\bf v})$ in the rest frame of the flow. The gravitational pull of the Sun deflects the particles, which speed up and get closer together as they approach the Sun. This so-called gravitational focusing effect changes the phase-space density of particles near the Sun.

Our goal is to compute the phase-space density of dark matter particles $f({\bf r}_E, {\bf v}_E)$ as a function of the particle velocity ${\bf v}_E$ at the position of Earth ${\bf r}_E$, starting from a given, but arbitrary, velocity distribution at infinity, i.e.\ far away from the Sun. 

The problem of the velocity distribution of a cloud of non-interacting particles in the gravitational field of a point mass  was discussed by Danby and Camm \cite{DC} many years ago. They gave an expression for the velocity distribution function at any point in the Sun's rest frame, but confined themselves to points along the axis of symmetry. Few years later, the same problem was considered again by Danby and Bray \cite{DB}, where they studied the gravitational enhancement of the density of a non zero temperature interstellar matter near a star by applying the velocity distribution function of Danby and Camm \cite{DC}. In their paper, they discussed the problem away from the axis of symmetry. In 1988 Griest \cite{G} studied the effect of the Sun's gravity on the distribution and detection of DM near the Earth in the case of the isothermal model. Griest \cite{G} considered the annual modulation for the signal of DM detection and obtained his velocity distribution function from Danby and Camm \cite{DC}, correcting however some errors and transforming it to the Earth's rest frame. He found that the inclusion of bound particles in calculating the distribution function has negligible effect on direct detection of DM. In 2002 Sikivie and Wick \cite{SW} analyzed the Sun's gravitational field for a flow with zero velocity dispersion through the Solar System. They gave expressions for the density and velocity distribution functions that they state were different from the analogous expressions in Danby and Camm \cite{DC} and Griest \cite{G}. In this paper we show that Danby and Camm's expression is correct provided their unspecified parameter $\beta$ is set to -1, that Danby and Bray's expression need corrections, that Griest's expression is correct after fixing a typo evident by comparing other equations in his paper, and that Sikivie and Wick's expression is correct. All expressions but Danby and Bray's match our analytic formulas and numerical results.

This paper is organized as follows. In Sec.~\ref{distribution function}, we give a general discussion of the distribution function for a flow of DM and its Boltzmann equation. In Sec.~\ref{THE ARRIVAL DISTRIBUTION: NUMERICAL METHOD}, we describe a numerical backward-in-time method to determine the DM particle distribution function based on an adjustable time-step and a predictor-corrector iteration. Using this method, in Sec.~\ref{Phase-space distribution} we show sky maps of the distribution of arrival directions of the particles on Earth. In Sec.~\ref{derivation of v}, an analytical expression for the particle velocity at infinity ${\bf v}_{\infty}$ is derived. When used to compute the particle distribution function, it gives the same results as in the numerical backward-in-time method. In Sec.~\ref{comparison with analytical expressions}, we compare our results to the four analytical expressions for unbound orbits mentioned above. Finally, we summarize our results in Sec.~\ref{conclusions}.

\section{The Distribution Function}
\label{distribution function}

In a flow of DM, particles move with a variety of velocities. Although it is hard to say what velocity an individual particle may have, it is possible to use a statistical approach to characterize the velocity attributes. The mathematical description of this statistical approach is called the phase-space distribution function $f({\bf r},{\bf v},t)$. 

For a system of $N$ identical classical particles, the distribution function $f({\bf r},{\bf v},t)$ is given by
\begin{equation}
dN = f({\bf r},{\bf v},t) \, d^3 x \, d^3 v ,
\end{equation}
where $dN$ is the number of particles in the volume element $d^3 x$ centered at ${\bf r}$ whose velocities fall in the velocity element $d^3 v$ centered at ${\bf v}$ at time $t$. From the distribution function one can compute the macroscopic variables for an ensemble of particles. For example, the density $\rho({\bf r},t)$, the number of particles per unit volume, is given by $\rho({\bf r},t) = \delta N / \delta V$, where $\delta N$ is the total number of particles in the volume element $\delta V$ at ${\bf r}$. In the limit $\delta V \to d^3 x$, $\delta N$ is the integral of $dN$ over all velocities. Hence 
\begin{equation}
\delta N = d^3 x \int f({\bf r},{\bf v},t) \, d^3 v
\end{equation}
and
\begin{equation}
\rho({\bf r},t) = \int f({\bf r},{\bf v},t) \, d^3 v .
\end{equation}

An equation that governs the evolution of the distribution function $f$ is the Boltzmann equation
\begin{equation}
\frac{\partial f}{\partial t} + v_{i} \frac{\partial f}{\partial x_{i}} + a_{i} \frac{\partial f}{\partial v_{i}} = \left(\frac{\partial f}{\partial t}\right)_{c} .
\end{equation}
Here $x_{i}$, $v_{i}$, and $a_{i}$ are the Cartesian components of the position, velocity, and acceleration vectors, respectively, and a summation is implied over repeated indices. The collision term on the right hand side receives contributions from particle collisions. However, cold dark matter is collisionless, thus 
\begin{equation}
\frac{\partial f}{\partial t} + v_{i} \frac{\partial f}{\partial x_{i}} + a_{i} \frac{\partial f}{\partial v_{i}} = 0 .
\end{equation}
For a stationary distribution function $f$ we have ${\partial f / \partial t} = 0$. In this case, the Boltzmann equation reduces to
\begin{equation}
v_{i} \frac{\partial f}{\partial x_{i}} + a_{i} \frac{\partial f}{\partial v_{i}} = 0 .
\end{equation}
It follows that $f$ is constant along trajectories, that is if ${\bf r} = {\bf r}(t)$ and ${\bf v} = {\bf v}(t)$ represent a particle trajectory, $f({\bf r}(t), {\bf v}(t)) = {\rm const}$, independent of $t$. This is Liouville's theorem.

To find the value of $f({\bf r}_E, {\bf v}_E)$ at the position of the Earth, we determine the velocity ${\bf v}_\infty $ it had when it was far away from the Sun. Then we use Liouville's theorem to equate $f({\bf r}_E, {\bf v}_E)$ to the distribution function at infinity $f_\infty({\bf v}_{\infty})$,
\begin{equation}
f({\bf r}_E, {\bf v}_E) = f_\infty({\bf v}_{\infty}) .
\label{ff1}
\end{equation}
This is the basis of our analysis.

It must be paid attention to the reference frame in which $f_{\infty}({\bf v}_\infty)$ is given. In our discussion, we assume it is {\it in the frame of the Sun}, namely that the velocity ${\bf v}_\infty$ is measured relative to the Sun.
For example, in the particle flow illustrated in Fig.~\ref{first}, which approaches the Sun with mean relative velocity ${\bf V}$ and with rest-frame distribution $f({\bf v})$, a Galilean transformation gives 
\begin{equation}
f_\infty({\bf v}_\infty) = f({\bf v}_\infty-{\bf V}) ,
\end{equation}
In particular, for a Maxwellian distribution, the distribution function in the rest frame of the flow
reads
\begin{equation}
f({\bf v}) = \frac{\rho_{\infty}}{(2 \pi \sigma^2)^{3/2}} \exp\!{\left(- {\frac{{\bf v}^{2}}{2 \sigma^2}}\right)} ,
\end{equation}
where $\rho_{\infty}$ is the uniform particle density at infinity and $\sigma$ is the velocity dispersion. In the Sun's frame, it becomes
\begin{equation}
f_{\infty}({\bf v}_{\infty}) = \frac{\rho_{\infty}}{(2 \pi \sigma^2)^{3/2}} \exp\!{\left(- {\frac{\left| {\bf v}_{\infty} - {\bf V} \right|^2}{2 \sigma^2}}\right)} .
\label{eq:f}
\end{equation}

\section{NUMERICAL METHOD}
\label{THE ARRIVAL DISTRIBUTION: NUMERICAL METHOD}

A backward-in-time method is used to compute the distribution function $f$ for a flow of DM particles at the Earth, given a velocity distribution $f_{\infty}({\bf v}_\infty)$ far away from the Sun.

\begin{figure}[t]
\centering
\includegraphics[width=5.5in]{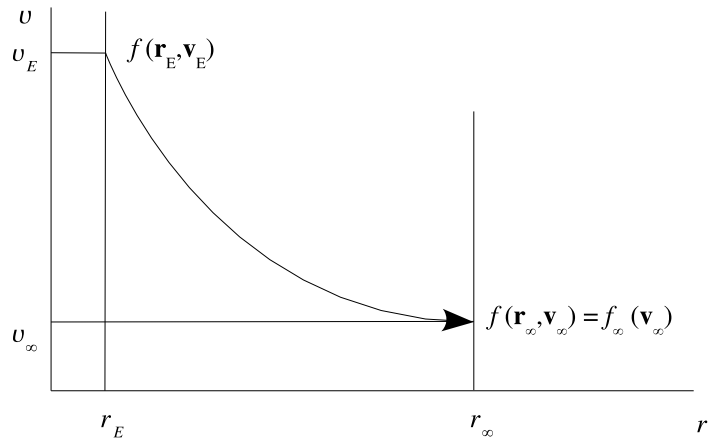}
\caption {The figure shows the idea of the backward-in-time method to compute the distribution function $f$.}
\label{time}
\end{figure}

To find the value of $f({\bf r}_E, {\bf v}_E)$ at the position of the Earth, we place a particle at the point $({\bf r}_E, {\bf v}_E)$, and follow its trajectory numerically backward in time until it is far away from the Sun, at position ${\bf r}_{\infty}$ with velocity ${\bf v}_{\infty}$ (see Fig.~\ref{time}). Since $f$ is constant along every trajectory (Liouville's theorem), we have 
\begin{equation}
f({\bf r}_E, {\bf v}_E) = f({\bf r}_{\infty}, {\bf v}_{\infty}) .
\label{ff}
\end{equation}
We set the latter equal to the given velocity distribution $f_{\infty}({\bf v}_\infty)$. We choose $\left | {\bf r}_{\infty} \right | = 90$ AU. 

The initial position ${\bf r}_0$ of each simulated trajectory is taken to coincide with the position of the Earth ${\bf r}_E$,
\begin{equation}
{\bf r}_0 = {\bf r}_E .
\end{equation}

The initial velocity ${\bf v}_0$ is set to the opposite of the arrival velocity of the particle on Earth,
\begin{equation}
{\bf v}_0 = - {\bf v}_E,
\end{equation}
so that the particle follows its trajectory backward in time.

For advancing particles along the trajectory, we use a modified version of the algorithm described in \cite{gravitybook}. Newton's theory of gravity is used to produce a simulation of the motion of particles in the gravitational field of the Sun. In the simulation, the trajectories of DM particles in the Solar System are traced as they are deflected by the Sun's gravitational field. The accuracy of the numerical calculation is maintained by two techniques: an adjustable time-step for advancing the orbit, and a predictor-corrector iteration for improving the accuracy of each step.

As is well-known, the acceleration of gravity on the particles due to the Sun (of mass $M$ and located at the origin of the coordinate system) is given by
\begin{equation}
{\bf a}({\bf r}) = - \frac{G M}{r^2} {\bf \hat r} ,
\label{a}
\end{equation}
where $G$ is Newton's gravitational constant.

To move the particle along the trajectory, the calculation uses a number of steps, up to 50,000, with a varying time-step $\Delta t$. If the acceleration were constant, equal to ${\bf a}$, say,  the velocity of the particle would change in a time $\Delta t$ by $\Delta {\bf v} = {\bf a} \Delta t$, and its position would change by $\Delta {\bf r} = \overline{\bf v} \Delta t$ with a velocity $\overline {\bf v}$ equal to  the average of the velocity ${\bf v}_{i}$ at the beginning of the step and the velocity ${\bf v}_{i+1}$ at the end. Since the  particle moves in a non-uniform gravitational field, there is an inevitable change in the acceleration. But if the change in position is sufficiently small, the error we make by assuming a constant acceleration can be made as small as we wish. 

Using the approximation that the position changes by the average velocity during the time-step and the velocity changes by the average acceleration, we have for the new velocity and position
\begin{eqnarray}
{\bf r}_{i+1} & = & {\bf r}_{i} + \frac{1}{2}({\bf v}_{i} + {\bf v}_{i+1})\Delta t ,
\label{dv}
\\
{\bf v}_{i+1} & = & {\bf v}_{i} + \frac{1}{2}({\bf a}_{i} + {\bf a}_{i+1})\Delta t .
\label{vi1}
\end{eqnarray}
where ${\bf a}_{i} = {\bf a}({\bf r}_{i}$), see Eq.~(\ref{a}).
This is a non-linear system of six equations in six unknowns, ${\bf r}_{i+1}$ and ${\bf v}_{i+1}$. 

To compute ${\bf v}_{i+1}$ from Eq.~(\ref{vi1}) we need to know the value of ${\bf r}_{i+1}$ that enters in ${\bf a}_{i+1}$, but to find ${\bf r}_{i+1}$ we need to know ${\bf v}_{i+1}$ in Eq.~(\ref{dv}). To break the cycle, we use an iterative method, i.e.\  a series ${\bf r}_{i+1}^{(k)}$, ${\bf v}_{i+1}^{(k)}$ of successively more accurate approximations to the  position and velocity at the end of the step (here $k$ is the iteration number). 
We start the iteration by setting
\begin{eqnarray}
{\bf r}^{(1)}_{i+1} & = & {\bf r}_{i} + {\bf v}_{i}\Delta t + \frac{1}{2} {\bf a}_{i} (\Delta t)^2 ,
\\
{\bf v}^{(1)}_{i+1} & = & {\bf v}_{i} + {\bf a}_{i} \Delta t .
\end{eqnarray}
We refine the estimate using Eqs.~(\ref{dv}) and~(\ref{vi1}) with ${\bf r}_{i+1}^{(k)}$ and ${\bf v}_{i+1}^{(k)}$ in the right hand side, and ${\bf r}_{i+1}^{(k+1)}$ and ${\bf v}_{i+1}^{(k+1)}$ in the left hand side. Our convergence condition is
\begin{equation}
\left| {\bf v}_{i+1}^{(k+1)} - {\bf v}_{i+1}^{(k)} \right| \leq \epsilon_{p} \left| {\bf a}_{i} \Delta t \right| ,
\end{equation}
where $\epsilon_{p}$ is a dimensionless number. After some trials, we found that $\epsilon_{p} = 10^{-3}$ was a good compromise between accuracy and efficiency.

The time-step is adjusted as follows. At every step $i$, we compute the acceleration ${\bf a}_{i+1}$ at the new position using a trial time-step, and then compare the change in acceleration ${\bf a}_{i+1} - {\bf a}_i$ during the step with the acceleration ${\bf a}_i$ at its beginning. If 
\begin{equation}
\left| {\bf a}_{i+1} - {\bf a}_{i} \right| > \epsilon_{tl} \left| {\bf a}_{i} \right| ,
\end{equation}
where $\epsilon_{tl}$ is a small positive number, we deem $\Delta t$ too large. In this case, we divide $\Delta t$ by two and restart the current step. Else, if
\begin{equation}
\left| {\bf a}_{i+1} - {\bf a}_{i} \right| \leq \epsilon_{ts} \left| {\bf a}_{i} \right| ,
\end{equation}
where $\epsilon_{ts}$ is another small positive number, we deem $\Delta t$ too small. In this case, we double $\Delta t$ and restart the current step. This is repeated until the time-step $\Delta t$ is acceptable. After some experimentation, we found that $\epsilon_{tl} = \epsilon_{ts} = 10^{-3}$ is a good choice.

\section{Analytical method}
\label{derivation of v}

As an alternative to the numerical method, and as a cross check of the calculation, the velocity at infinity ${\bf v}_\infty$ appearing in Eq.~(\ref{ff1}) can also be computed analytically.

The expression for ${\bf v}_{\infty}$ can be obtained using the conservation of the Laplace-Runge-Lenz vector following Sikivie and Wick \cite{SW}. We give here a shorter derivation that will lead to a simpler formula for ${\bf v}_{\infty}$.

\begin{figure}[t]
\centering
\includegraphics[width=4.5in]{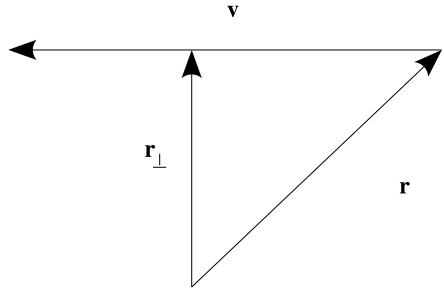}
\caption {Definition of ${\bf r}_{\perp}$ as used in the derivation of ${\bf v}_{\infty}$.}
\label{figvr}
\end{figure}

The Laplace-Runge-Lenz vector per unit mass is 
\begin{equation}
{\bf A} = {\bf v} \times ({\bf r} \times {\bf v}) - G M {\bf \hat r} = r_{\perp} v^2 {\bf \hat r}_{\perp} - G M {\bf \hat r} .
\end{equation}
Here ${\bf \hat r}_{\perp}= {\bf r}_{\perp} / r_{\perp}$, and ${\bf r}_\perp$ is the projection of ${\bf r}$ perpendicular to ${\bf v}$ (see Fig.~\ref{figvr}).

\begin{figure}[t]
\centering
\includegraphics[width=5.5in]{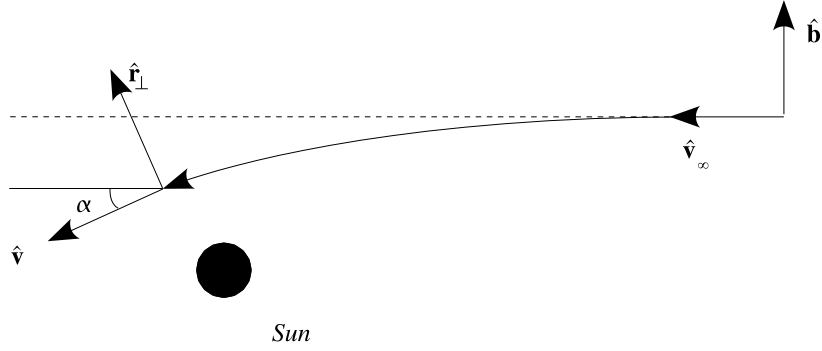}
\caption {Rotation of the basis vectors as a particle moves in the Sun's gravitational field.}
\label{figrot}
\end{figure}

Conservation of the Laplace-Runge-Lenz vector ${\bf A}$ implies
\begin{equation}
r_{\perp} v^2 {\bf \hat r}_{\perp} - G M {\bf \hat r} = b v_{\infty}^2 {\bf \hat b} + G M {\bf \hat v}_{\infty} ,
\label{eq:A}
\end{equation}
where $b$ is the impact parameter and ${\bf\hat b}$ is a unit vector perpendicular to ${\bf v}_\infty$ in the same plane as ${\bf v}$ and ${\bf r}$ (see Fig.~\ref{figrot}).

Conservation of angular momentum in the form
\begin{equation}
b v_{\infty} = r_{\perp} v 
\end{equation}
allows us to eliminate $b$ from Eq.~(\ref{eq:A}). Dividing the result by $r$ and using the identity ${\bf r} = {\bf r}_{\perp} + {\bf r} \cdot {\bf \hat v} {\bf \hat v}$, the conservation of ${\bf A}$ equation becomes
\begin{equation}
\frac{r_{\perp}}{r} \left(v^2 - \frac{G M}{r}\right) {\bf \hat r}_{\perp} - \frac{G M}{r} {\bf \hat r} \cdot {\bf \hat v} {\bf \hat v} = \frac{r_{\perp}}{r} v v_{\infty} {\bf \hat b} + \frac{G M}{r} {\bf \hat v}_{\infty} .
\label{express}
\end{equation}

Eq.~(\ref{express}) is a relation between the unit vector bases $( {\bf \hat r}_\perp, {\bf\hat v})$ and $( {\bf \hat b}, {\bf\hat v}_\infty)$ (see Fig.~\ref{figrot}). Introducing the rotation angle $\alpha$ between the two bases, we write
\begin{eqnarray}
{\bf \hat b} & = & \cos\alpha {\bf \hat r}_{\perp} - \sin\alpha {\bf \hat v} ,
\label{bhat}
\\
{\bf \hat v}_{\infty} & = & \sin\alpha {\bf \hat r}_{\perp} + \cos\alpha {\bf \hat v} ,
\label{vhat}
\end{eqnarray}

Inserting Eqs.~(\ref{bhat}) and (\ref{vhat}) into Eq.~(\ref{express}), we have, in the $( {\bf \hat r}_\perp, {\bf\hat v})$ basis,
\begin{eqnarray}
\frac{r_{\perp}}{r} v v_{\infty} \cos\alpha + \frac{G M}{r} \sin\alpha & = & \frac{r_{\perp}}{r} \left(v^2 - \frac{G M}{r} \right) ,
\\
- \frac{G M}{r} \cos\alpha + \frac{r_{\perp}}{r} v v_{\infty} \sin\alpha  & = & \frac{G M}{r} {\bf \hat r} \cdot {\bf \hat v} .
\end{eqnarray}
Solving this linear system for $\cos \alpha$ and $\sin\alpha$, we find
\begin{eqnarray}
\cos\alpha & = & \frac{1}{D} \left[\frac{r_{\perp}^2}{r^2} v v_{\infty} \left(v^2 - \frac{G M}{r}\right) - \left(\frac{G M}{r}\right)^2 {\bf \hat r} \cdot {\bf \hat v}\right] ,
\\
\sin\alpha & = & \frac{1}{D} \frac{r_{\perp}}{r} \frac{G M}{r} \left(v^2 - \frac{G M}{r} + v v_{\infty} {\bf \hat v} \cdot {\bf \hat r}\right) ,
\end{eqnarray}
where
\begin{equation}
D = \left(v^2 - \frac{G M}{r} + v v_{\infty} {\bf \hat v} \cdot {\bf \hat r}\right) \, \left(v^2 - \frac{G M}{r} - v v_{\infty} {\bf \hat v} \cdot {\bf \hat r}\right) .
\label{A2}
\end{equation}
Energy conservation in the form
\begin{equation}
v_{\infty} = \sqrt{v^2 - \frac{2 G M}{r}} ,
\label{vinfty2}
\end{equation}
has been used in deriving the above expression for $D$. Notice in passing that $D = |{\bf A}|^2/r^2$.

Substituting the above expressions for $\cos\alpha$ and $\sin\alpha$ into Eq.~(\ref{vhat}), multiplying by $v_\infty$, and canceling common factors, we finally obtain the equation
\begin{equation}
{\bf v}_{\infty} = \frac{v_{\infty}^2 {\bf v} + v_\infty (GM/r) {\bf \hat r} - v_\infty {\bf v} ({\bf v} \cdot {\bf \hat r})}{ v_{\infty}^2 + (GM/r) - v_\infty ({\bf v} \cdot {\bf \hat r}) } 
\label{v}
\end{equation}
where $v_{\infty}$ is given in Eq.~(\ref{vinfty2}). As a simple check, when $M = 0$, Eq.~(\ref{v}) correctly gives ${\bf v}_{\infty} = {\bf v}$. In our application, ${\bf v}$ and ${\bf r}$ are the particle velocity and position when it reaches the Earth, namely ${\bf v} = {\bf v}_E$ and ${\bf r} = {\bf r}_E$. 

The expression for ${\bf v}_{\infty}$ in Eq.~(\ref{v}) depends on the Keplerian character of the potential, but is independent of the velocity distribution function at infinity. This means that it can be used not only for a Maxwellian distribution but also for more general functions. 

When we used Eq.~(\ref{v}) in the velocity distribution function, Eq.~(\ref{eq:f}), we achieved an excellent agreement with our numerical backward-in-time method described in the previous section. Later, in Sec.~\ref{comparison with analytical expressions}, we will show that the expression for ${\bf v}_{\infty}$ in Eq.~(\ref{v}) matches formulas used by Griest \cite{G} and Sikivie and Wick \cite{SW}.

\section{Phase-space distribution}
\label{Phase-space distribution}

In this section, we present the velocity distribution of the DM particles as they arrive at the Earth. We place the Earth at four different locations during the year, and plot the flux of WIMPs as a function of their arrival direction ${\bf\hat n}$ at fixed arrival speed $v_E$.

\subsection{Positions of the Earth}

\begin{figure}[t]
\centering
\includegraphics[width=5.0in]{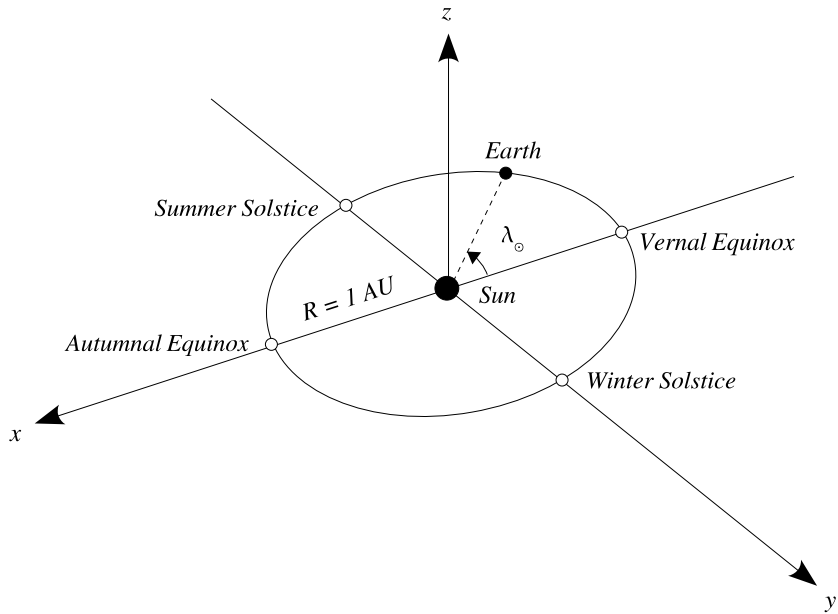}
\caption {The ecliptic longitude of the Sun $\lambda_{\odot}$ changes from $0^{\circ}$ to $360^{\circ}$ as the Earth orbits the Sun.}
\label{earthsun}
\end{figure}

We specify the position of the Earth at different times of the year by means of the ecliptic longitude of the Sun $\lambda_{\odot}$, which varies from $0^\circ$ to $360^\circ$ during the course of one year. Fig.~\ref{earthsun} shows how $\lambda_{\odot}$ changes as the Earth orbits the Sun. At the Vernal (or Spring) Equinox ($\approx$ March 21), $\lambda_\odot = 0^\circ$; at the Summer Solstice ($\approx$ June 20), $\lambda_\odot = 90^\circ$; at the Autumnal Equinox ($\approx$ September 21), $\lambda_\odot = 180^\circ$; and at the Winter Solstice ($\approx$ December 20), $\lambda_\odot = 270^\circ$.  Table \ref{table seasons} summarizes the relation between the ecliptic coordinates of the Sun, the beginning of the four astronomical seasons, and their approximate calendar dates.

\begin{table}[bp]
\caption{The table relates the position of the Sun in the sky in ecliptic coordinates ($\lambda_{\odot}, \beta_\odot)$ to the beginning of the four astronomical seasons and their approximate dates.}
\begin{tabular}{|c|l|l|}
\hline\
Sun's ecliptic coordinates $(\lambda_{\odot},\beta_\odot)$ &  Season & Date \\
\hline
$(0^{\circ},0^{\circ})$ & Vernal Equinox & $\approx$ March 21 \\
$(90^{\circ},0^{\circ})$ & Summer Solstice & $\approx$ June 20 \\
$(180^{\circ},0^{\circ})$ & Autumnal Equinox & $\approx$ September 21 \\
$(270^{\circ},0^{\circ})$ & Winter Solstice & $\approx$ December 20 \\
\hline
\end{tabular}
\label{table seasons}
\end{table}

In the approximation of a circular orbit, one has the following relation between $\lambda_\odot$ and the time of year $t$:
\begin{equation}
\lambda_{\odot} \simeq \frac{360^\circ}{1 \rm yr}(t - t_{\rm VE}) .
\end{equation}
Here $t$ is the time during the year, which e.g.\ runs from $0$ at New Year's midnight to $1 {\rm yr}\simeq 365.2425$ days at the end of the year. The time $t_{\rm VE}$ is the time of the Vernal Equinox, i.e.\  $t_{\rm VE} \simeq 79.25$ days from New Year's midnight to the midnight of March 21.  

We fix the origin of our coordinate system at the position of the Sun (see Fig.~\ref{earthsun}). The $x$ axis points in the direction of the Earth's position at the Autumnal Equinox, the $y$ axis in the direction of the Earth's position at the Winter Solstice, and the $z$ axis is perpendicular to the Earth's orbit. As seen from the Earth, the $x$ axis points toward the position of the Sun at the Vernal Equinox, the $y$ axis toward the position of the Sun at the Summer Solstice, and the $z$ axis toward the North Pole of the ecliptic.

The position vector of the Earth is given by
(for a circular orbit)
\begin{equation}
{\bf r}_E = - \left( \cos\lambda_{\odot} {\bf\hat x} + \sin\lambda_{\odot} {\bf\hat y} \right) {\rm AU}.
\end{equation}
Here 1 AU is the average distance between the Earth and the Sun.

\subsection{Velocities of the DM particles}

To ensure that only unbound orbits are considered, we constrain the particle speed $v_E$ to be equal to or greater than the escape speed $v_{\rm esc}$ at the Earth's position,
\begin{equation}
v_E \ge v_{\rm esc} = \sqrt{\frac{2 G M}{r_E}} .
\end{equation}

\begin{figure}[t]
\centering
\includegraphics[width=4.5in]{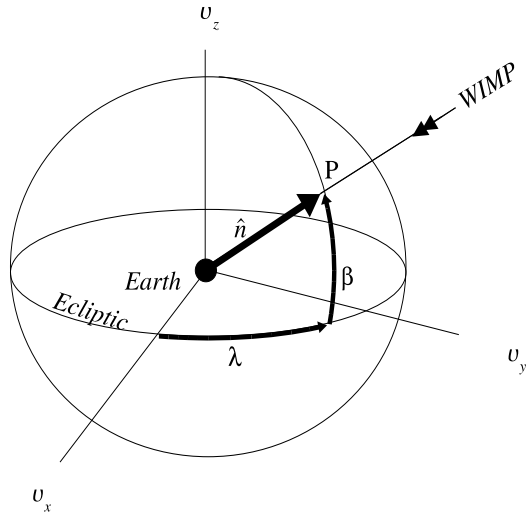}
\caption {Specification of the arrival direction of a dark matter particle by means of a unit vector ${\bf \hat n}$ and ecliptic coordinates $\lambda$ and $\beta$ on the celestial sphere.}
\label{unitsphere}
\end{figure}

We specify the arrival direction of DM particles using a unit vector ${\bf \hat n}$ pointing in the direction of arrival. We represent the sky by a sphere centered at the Earth (celestial sphere, see Fig.~\ref{unitsphere}).  We use the ecliptic coordinate system in which the two coordinates of a point P on the celestial sphere are: (1) the ecliptic latitude $\beta$, which is the angular distance from the ecliptic to P and varies from $+90^{\circ}$ at the Ecliptic North Pole to $-90^{\circ}$ at the Ecliptic South Pole, and (2) the ecliptic longitude $\lambda$, which is the angular distance along the ecliptic from the Vernal equinox to the great circle through P and is measured eastwards in degrees from $0^{\circ}$ to $360^{\circ}$.

In ecliptic coordinates, 
\begin{equation}
{\bf \hat n}(\lambda,\beta) = \cos\beta \cos\lambda {\bf\hat x} + \cos\beta \sin\lambda {\bf\hat y} + \sin\beta {\bf\hat z}.
\end{equation}

As can be seen from Fig.~\ref{unitsphere}, since the unit vector ${\bf\hat n}$ points in the direction {\it from which} the DM particle is coming, its velocity points in the opposite direction. Thus the velocity of the particle at the Earth is
\begin{equation}
{\bf v}_E = - v_E {\bf\hat n} = - v_E (\cos\beta \cos\lambda {\bf \hat x} + \cos\beta \sin\lambda {\bf \hat y} + \sin\beta {\bf \hat z}) ,
\label{eq:v0}
\end{equation}

In the backward-in-time method, the initial velocity of the DM particle used to start the simulation is opposite to the actual velocity of the particle at the Earth, that is ${\bf v}_0 = -{\bf v}_E = + v_E {\bf \hat n}$.

\subsection{Sky maps of the DM particle flux}
\label{distribution of WIMPs on the sky}

In this Section, we present sky maps that show the arrival distribution of the DM particles for three cases: a stream moving in the $+y$ direction, a stream moving in the $-x$ direction, and the Maxwellian distribution of the standard halo.

We start by deriving a formula for the flux of DM particles that reach the Earth. The number of particles at the Earth position is
\begin{equation}
d N = f({\bf r}_E, {\bf v}_E) \, d^3 x \, d^3 v .
\label{dn}
\end{equation}
For a flux of particles, we write the volume element as
\begin{equation}
d^3 x = dA \, dl = dA \, v_E \, dt ,
\label{d3x}
\end{equation}
where $dA$ is an element of area orthogonal to the particle flux, and $dl = v_E dt$ is the length traveled by the particles of speed $v_E$ in a time $dt$. We use spherical coordinates in velocity space, and write the velocity element as
\begin{equation}
d^3 v = v_E^2 \, dv_E \, d\Omega ,
\label{d3v}
\end{equation}
where $d\Omega$ is the solid angle in velocity space covered by the arrival directions of the particles. Inserting Eqs.~(\ref{d3x}) and (\ref{d3v}) into Eq.~(\ref{dn}), the number of particles becomes
\begin{equation}
d N = f({\bf r}_E, {\bf v}_E) \, v_E^3 \, dA \, dt \, dv_E \, d\Omega .
\end{equation}
We define the specific flux of DM particles reaching the Earth from a direction ${\bf \hat n} = - {\bf\hat v}_E$ as the flux of particles per unit area per unit solid angle per unit speed
\begin{equation}
F(v_E,{\bf\hat n}) \equiv \frac{dN}{dA \, dt \, dv_E \, d\Omega} = f({\bf r}_E, {\bf v}_E) \, v_E^3 .
\label{eq:N}
\end{equation}

We produce sky maps of $F(v_E, {\bf\hat n})$ by fixing the value of $v_E$ and varying ${\bf\hat n}$ on a regular grid of 360 points in $\lambda$ and 100 points in $\sin\beta$. A gray scale is used to represent the values of $F(v_E, {\bf\hat n})/\rho_\infty$, darker gray levels corresponding to lower values of the specific flux. In all the maps, the observer is on the Earth and looking toward the $+x$ axis (see Fig.~\ref{earthsun}), and the arrival speed of the DM particles  is fixed at $v_E = 30$ AU/yr.

\begin{figure}[tbp]
\centering
\includegraphics[width=6.0in]{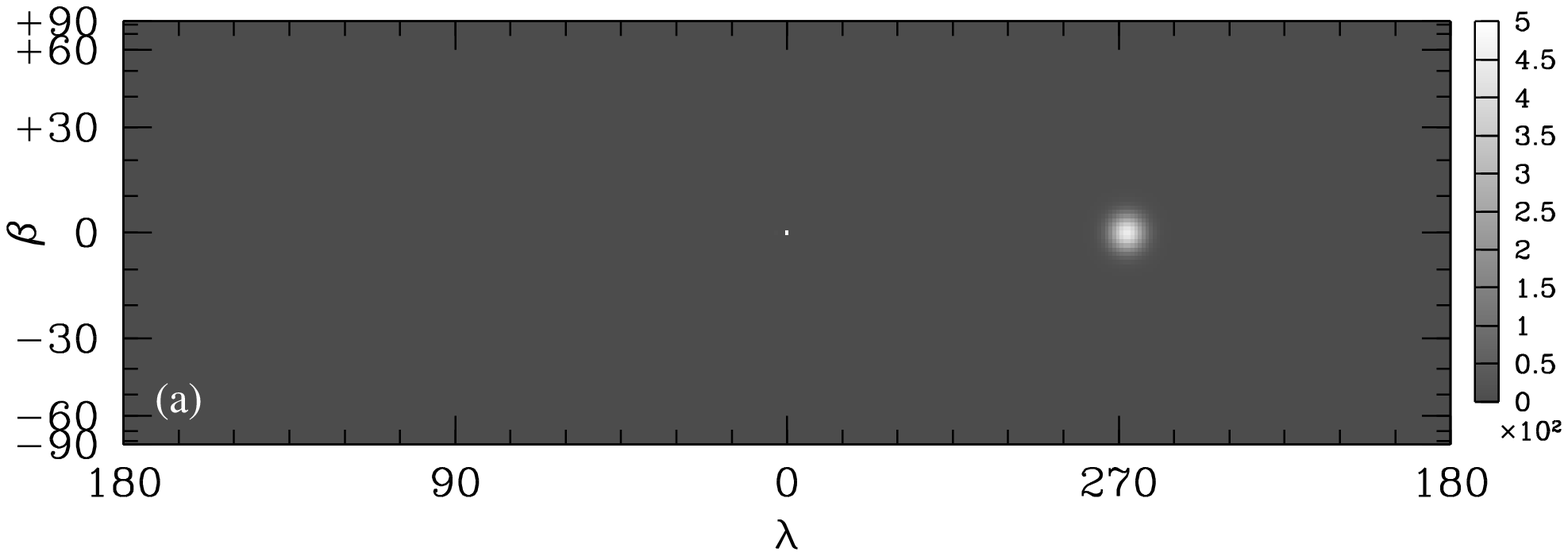}
\includegraphics[width=6.0in]{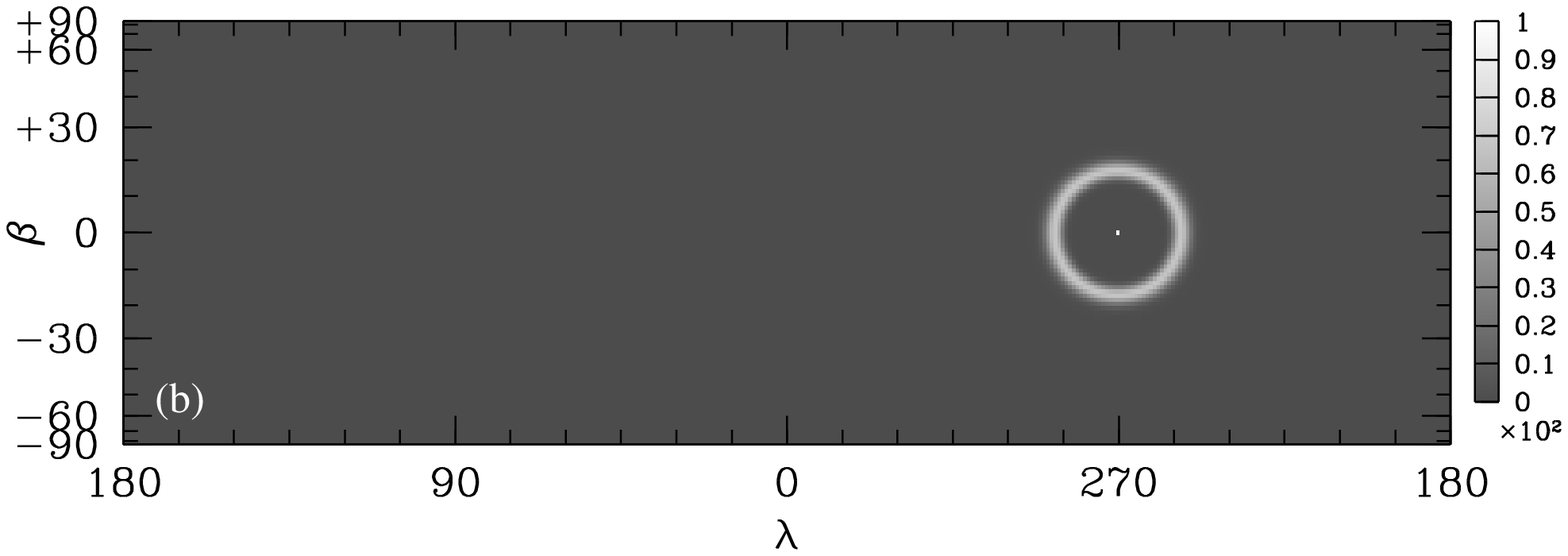}
\caption {Specific flux of dark matter particles $F(v_E,{\bf\hat n})$ for a stream moving toward the $+y$ direction at (a) Vernal Equinox, (b) Winter Solstice. The gray scale on the right shows the values of $F(v_E,{\bf\hat n})/\rho_\infty$.}
\label{fig7}
\end{figure}

Fig.~\ref{fig7} shows the specific flux for a stream of DM particles moving toward the $+y$ direction with velocity $V_y=10$ AU/yr and $V_x = V_z = 0$. The stream velocity distribution at infinity is assumed to be gaussian, Eq.~(\ref{eq:f}), with velocity dispersion $\sigma = 1$ AU/yr. This value of $\sigma$ is chosen so as to reduce the dispersion of the flow and have a clear and sharp view of the distribution. 
The map in Fig.~\ref{fig7}(a) is for $\lambda_{\odot} = 0^{\circ}$, which occurs at the time of the Vernal Equinox. In this map, the Sun is represented by a white dot at $(\lambda,\beta)=(0^{\circ},0^{\circ})$. The specific flux is concentrated in the circular spot centered at $(\lambda,\beta)=(270^{\circ},0^{\circ})$, which is in the $-y$ direction, as it should. The map in Fig.~\ref{fig7}(b) is for $\lambda_{\odot} = 270^{\circ}$, which occurs at the time of the Winter Solstice. In this map, the specific flux has spread out into a ring around the Sun. The origin of this ring can be understood by referring to Fig.~\ref{first}. In the case we are discussing, the Earth is located at the point on the left of the Sun where the two trajectories drawn cross each other. An observer on the Earth sees particles coming from two directions in the plane of the figure. One is the flux of particles passing on one side of the Sun, the other is the flux of particles passing on the other side. When the figure is rotated around the Earth-Sun axis, the two directions trace the ring in Fig.~\ref{fig7}(b). This ring is analogous to the Einstein ring in the gravitational lensing of light. 

\begin{figure}[tbp]
\centering
\includegraphics[width=6.0in]{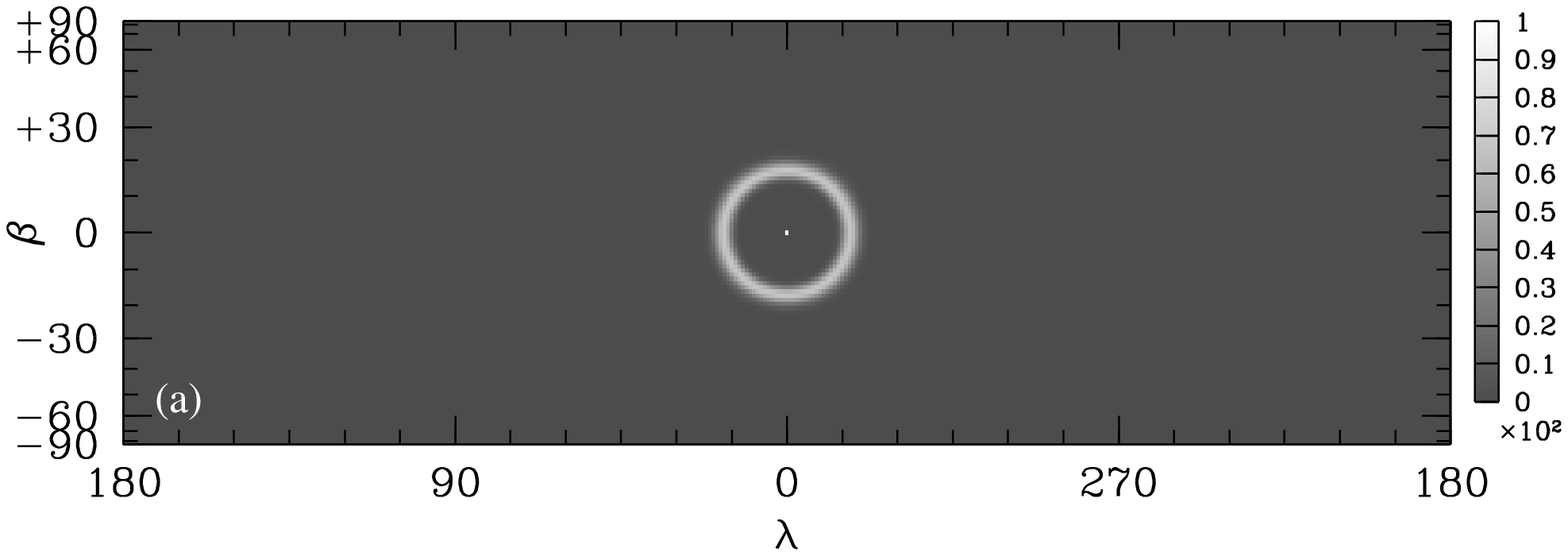}
\includegraphics[width=6.0in]{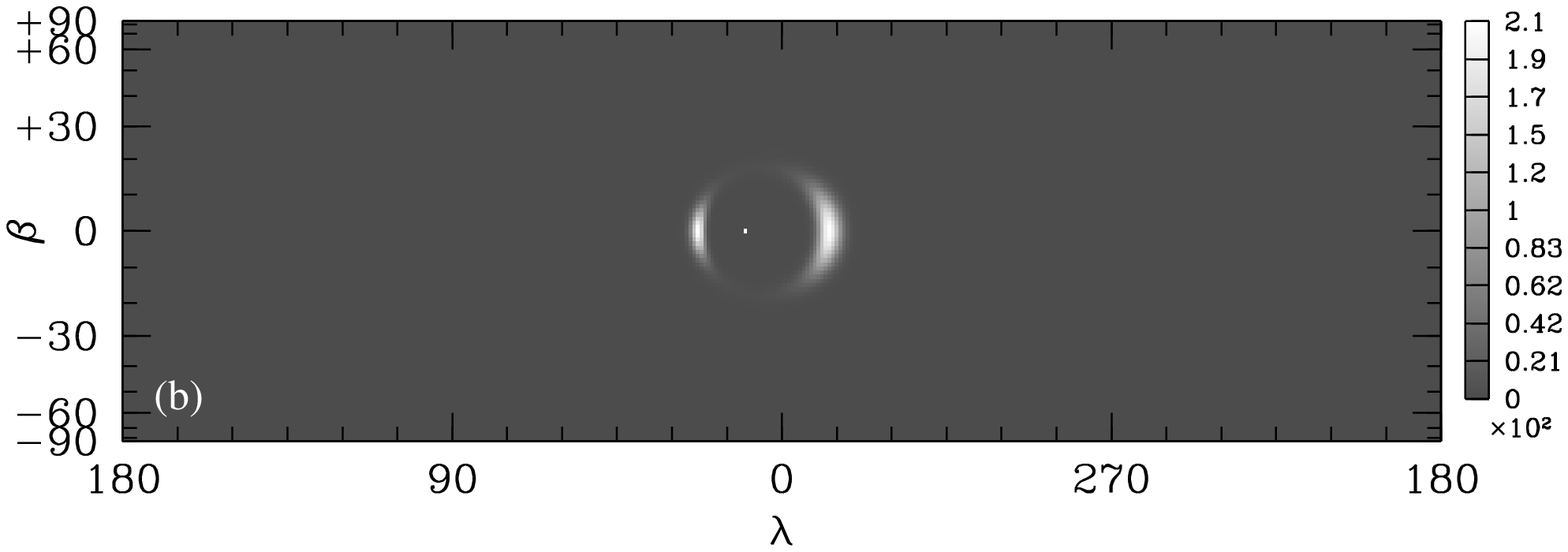}
\includegraphics[width=6.0in]{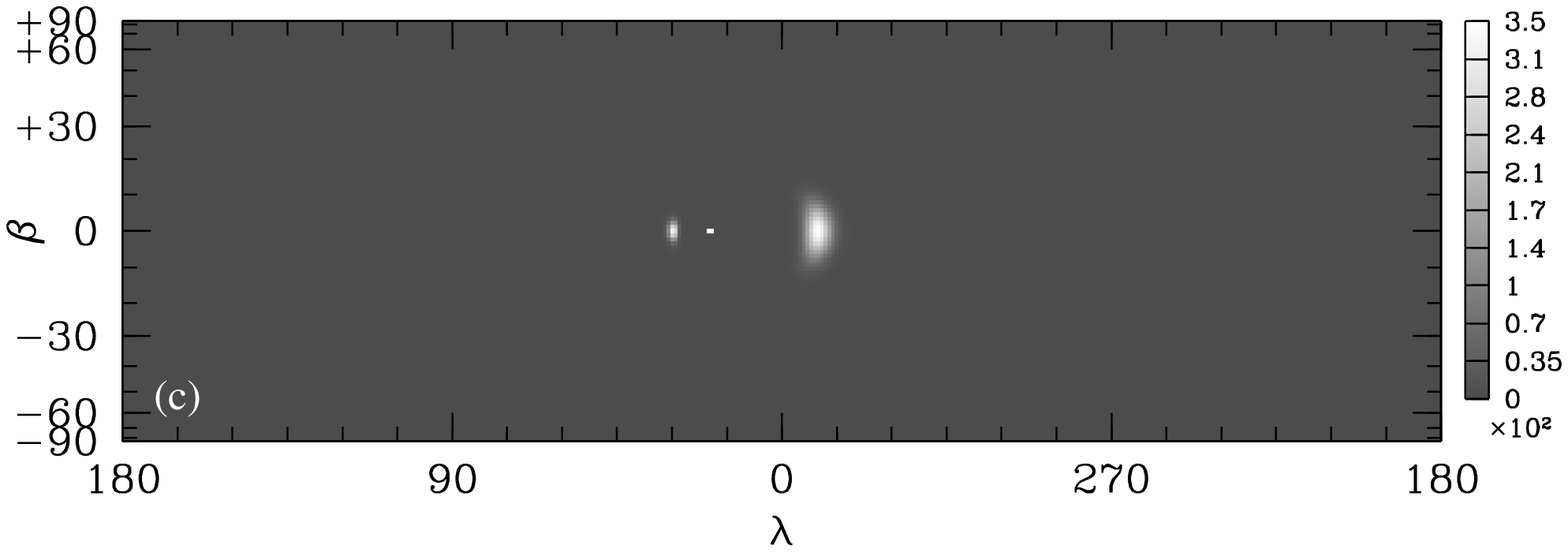}
\includegraphics[width=6.0in]{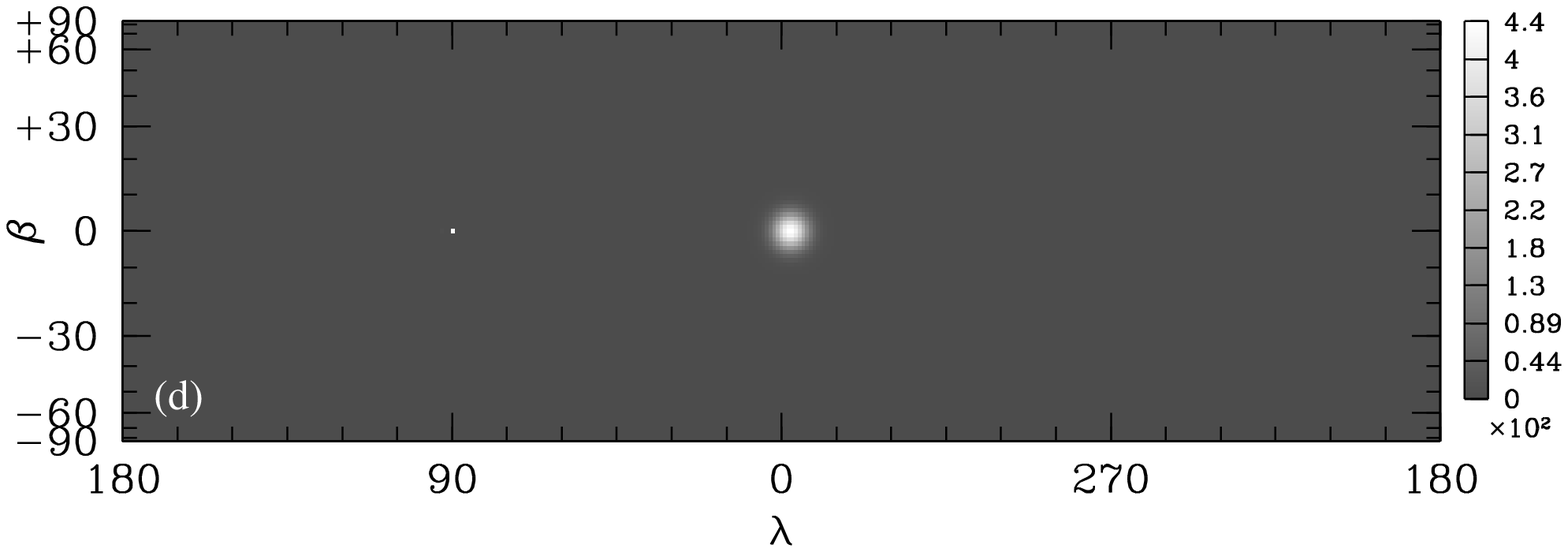}
\caption {Sky maps showing the arrival distribution of 30 AU/yr dark matter particles moving in the $-x$ with velocity dispersion 1 AU/yr. (a) $\lambda_{\odot} =0^{\circ}$ (Vernal Equinox), (b) $\lambda_\odot \simeq 9\fdg9$, (c) $\lambda_\odot = 19\fdg7$, and (d) $\lambda_\odot=90^\circ$ (Summer Solstice).}
\label{vxphi}
\end{figure}

As a second example, we consider a stream approaching the Solar System from the $+x$ direction, with
$V_x = -10$ AU/yr and $V_y = V_z = 0$. We again assume a Maxwellian velocity distribution at infinity with dispersion $\sigma=1$ AU/yr. The corresponding particle specific flux is shown in Fig.~\ref{vxphi} at four successive times between the Vernal Equinox ($\approx$ March 21) and the Summer Solstice ($\approx$ June 20). At the Vernal Equinox, Fig.~\ref{vxphi}(a), the Sun is directly in front of the stream, at $\lambda_\odot = 0^{\circ}$. The Earth is on the leeward side of the stream, i.e.\ directly behind the Sun (stream, Sun, and observer are lined up in this order). In this case, the observer sees the distribution of particles as a ring around the Sun, as explained before for Fig.~\ref{fig7}(b). Ten days later, Fig.~\ref{vxphi}(b), the Sun has shifted eastward to $\lambda_\odot \simeq 9\fdg9$. Now the observer sees an incomplete ring, because the Earth has moved to a point where fewer particle trajectories cross each other. As the observer's location changes again (see Fig.~\ref{vxphi}(c) for $\lambda_\odot \simeq 19\fdg7$, 20 days after the Vernal Equinox), the ring starts to disappear because fewer and fewer trajectories cross. Eventually, the observer sees a circular spot, like in Fig.~\ref{vxphi}(d) at $\lambda_\odot = 90^{\circ}$, three months after the Vernal Equinox. 

Finally, we give an example of arrival distributions under the assumption of a standard dark halo. In this model, the DM particles are on average at rest relative to the Galaxy, and their velocity distribution is Maxwellian with velocity dispersion $\sigma = v_{\rm LSR} / \sqrt{2}$. Here $v_{\rm LSR}$ is the speed of the Local Standard of Rest (LSR), which moves at $v_{\rm LSR} = 220$ km/s relative to the Galactic rest frame in the direction of the Galactic rotation. The latter is $(l,b) = (90^{\circ},0^{\circ})$ in Galactic coordinates, and $(\lambda,\beta) = ({347\fdg340},{59\fdg574})$ in ecliptic coordinates. In Astronomical Units, $\sigma = 32.816$ AU/yr. The other parameter we need is the mean velocity of the flow ${\bf V}$ with respect to the Sun. Alternatively, we can specify its opposite $-{\bf V}$, i.e. the velocity of the Sun with respect to the flow. In the standard halo, the DM particles are on average at rest in the Galactic rest frame, thus $-{\bf V}$ is the velocity of the Sun with respect to the Galactic rest frame. We write it as the sum of the velocity ${\bf v}_{\rm Sun-LSR}$ of the Sun with respect to the LSR (called the solar motion) and the velocity ${\bf v}_{\rm LSR}$ of the LSR with respect to the Galactic rest frame. We take the galactic components of the solar motion to be \cite{WJ} $U = 10.00 \pm 0.36$ km/s (radially inwards), $V = 5.25 \pm 0.62$ km/s (in the direction of Galactic rotation) and $W = 7.17 \pm 0.38$ km/s (vertically upwards). The central values have a magnitude $v_{\rm Sun-LSR} = 13.378$ km/s $ = 2.822$ AU/yr in the direction $(l,b) = (27\fdg70, 32\fdg409)$ or $(\lambda,\beta) = (248\fdg35,32\fdg189)$. 
Putting things together, we have
\begin{equation}
{\bf V} = - {\bf v}_{\rm LSR} - {\bf v}_{\rm Sun-LSR} .
\end{equation}
Converting from the Galactic to the ecliptic coordinate system and then our ($x,y,z$) coordinate system, we obtain ${\bf V} = (-22.049, 7.372, -41.521)$ AU/yr in the direction $(\lambda,\beta) = (-18\fdg487,60\fdg755)$. 

\begin{figure}[tbp]
\centering
\includegraphics[width=6.0in]{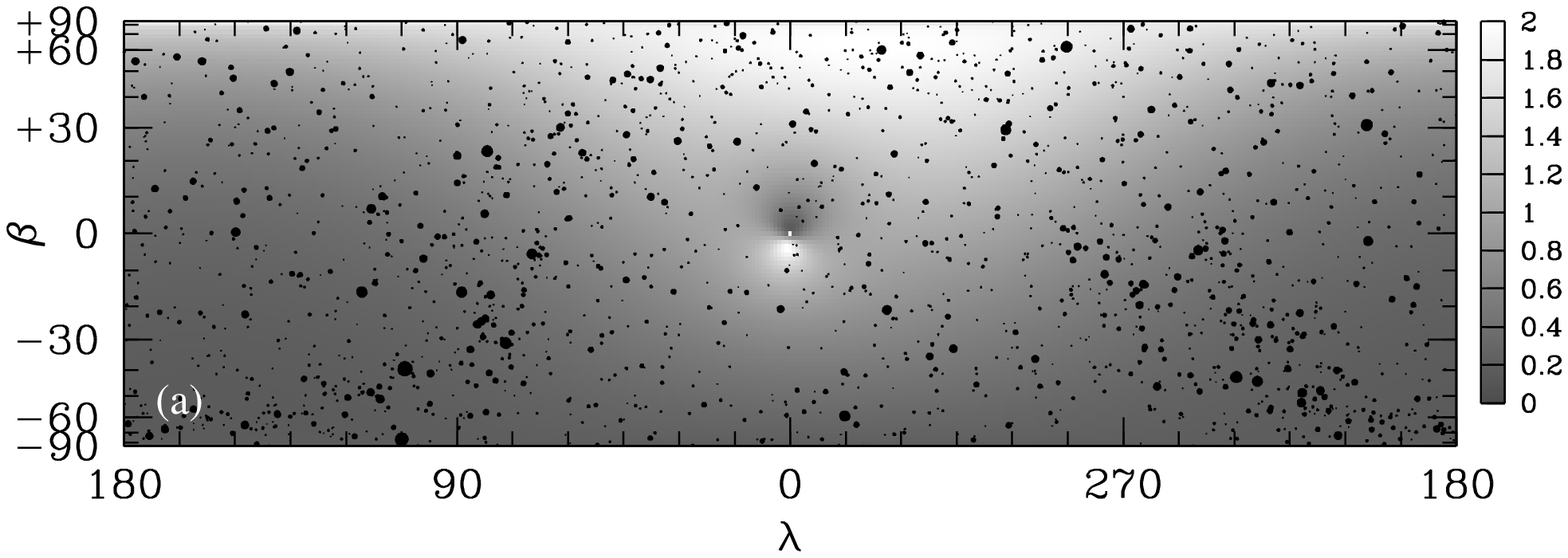}
\includegraphics[width=6.0in]{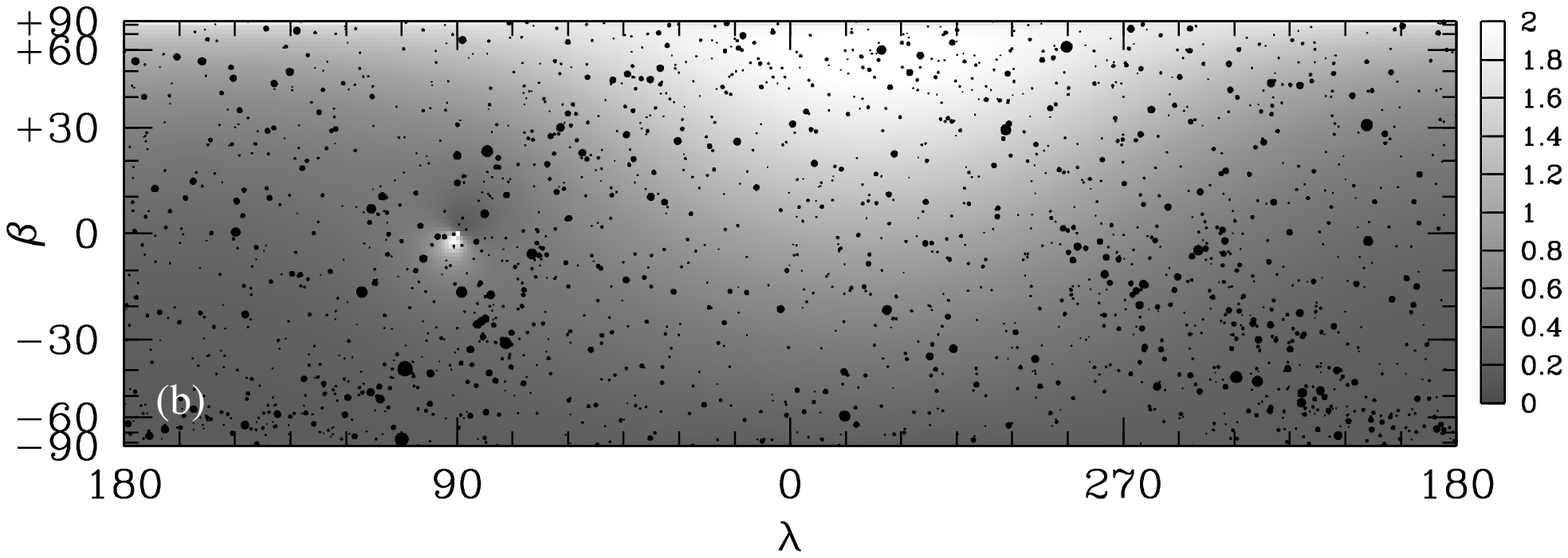}
\includegraphics[width=6.0in]{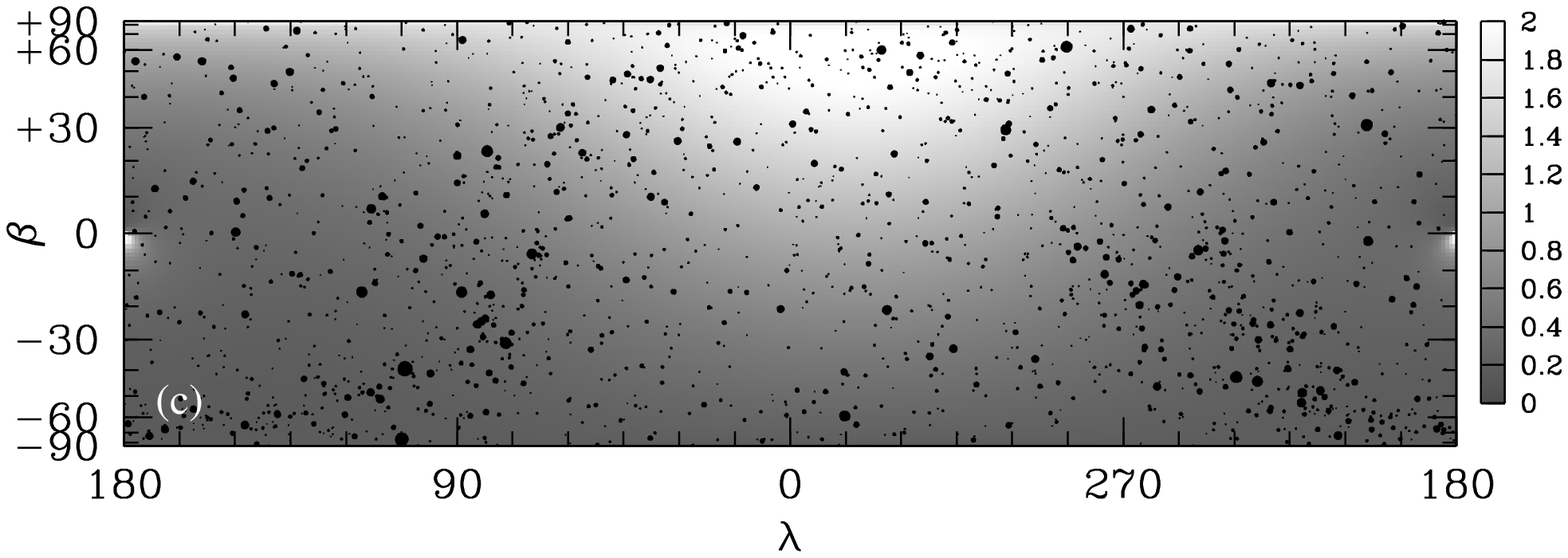}
\includegraphics[width=6.0in]{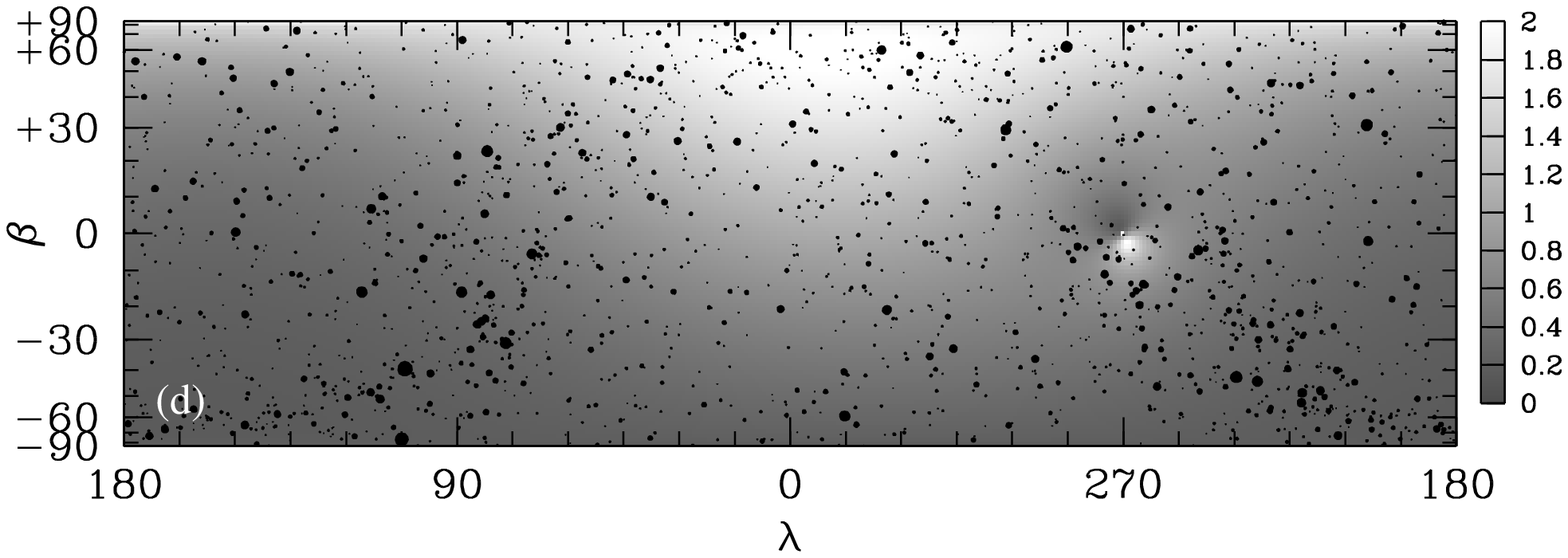}
\caption {Sky maps showing the arrival distribution of 30 AU/yr DM particles using standard halo parameters for $\sigma$ and ${\bf V}$. (a) $\lambda_{\odot} =0^{\circ}$ (Vernal Equinox), (b) $\lambda_{\odot} =90^{\circ}$ (Summer Solstice), (c) $\lambda_{\odot} =180^{\circ}$ (Autumnal Equinox), and (d) $\lambda_{\odot} =270^{\circ}$ (Winter Solstice).}
\label{par}
\end{figure}

The arrival distribution of DM particles for $v_E = 30$ AU/yr at four times of the year is shown in Fig.~\ref{par}. As before, the intensity of the gray levels represents the values of the particle specific flux $F(v_E,{\bf\hat n})/\rho_\infty$, with whiter regions corresponding to a higher flux. The effect of gravitational focusing near the Sun is evident. In order to help the reader place the DM particle distribution on the sky, the location of the $1986$ brightest stars (brighter than $\sim 5.17$ visual magnitude) is indicated by black dots. The size of each dot is proportional to the star's magnitude. Notice the constellation of Orion at $\lambda \approx 85^{\circ}$, $\beta \approx -30^{\circ}$, and the constellation of Ursa Major at the top left of the figure. An almost sinusoidal ``band'' of stars can be followed from the lower left corner through Orion up to the top center and symmetrically down: it is the Milky Way.

\section{COMPARISON WITH previous analyses}
\label{comparison with analytical expressions}

In this section, we compare our results with analytical expressions of Danby and Camm \cite{DC}, Danby and Bray \cite{DB}, Griest \cite{G}, and Sikivie and Wick \cite{SW}. We find Danby and Camm's expression correct after fixing their unspecified parameter $\beta$, Danby and Bray's expression incorrect unless few signs are changed, Griest's expression correct after fixing a typo, and Sikivie and Wick's expression correct. We checked that all distributions, with the exception of Danby and Bray's, match the results of our numerical and analytical methods shown in Figs.~\ref{fig7}, \ref{vxphi}, and~\ref{par}.

\subsection{Danby and Camm}

\begin{figure}[t]
\centering
\includegraphics[width=5.0in]{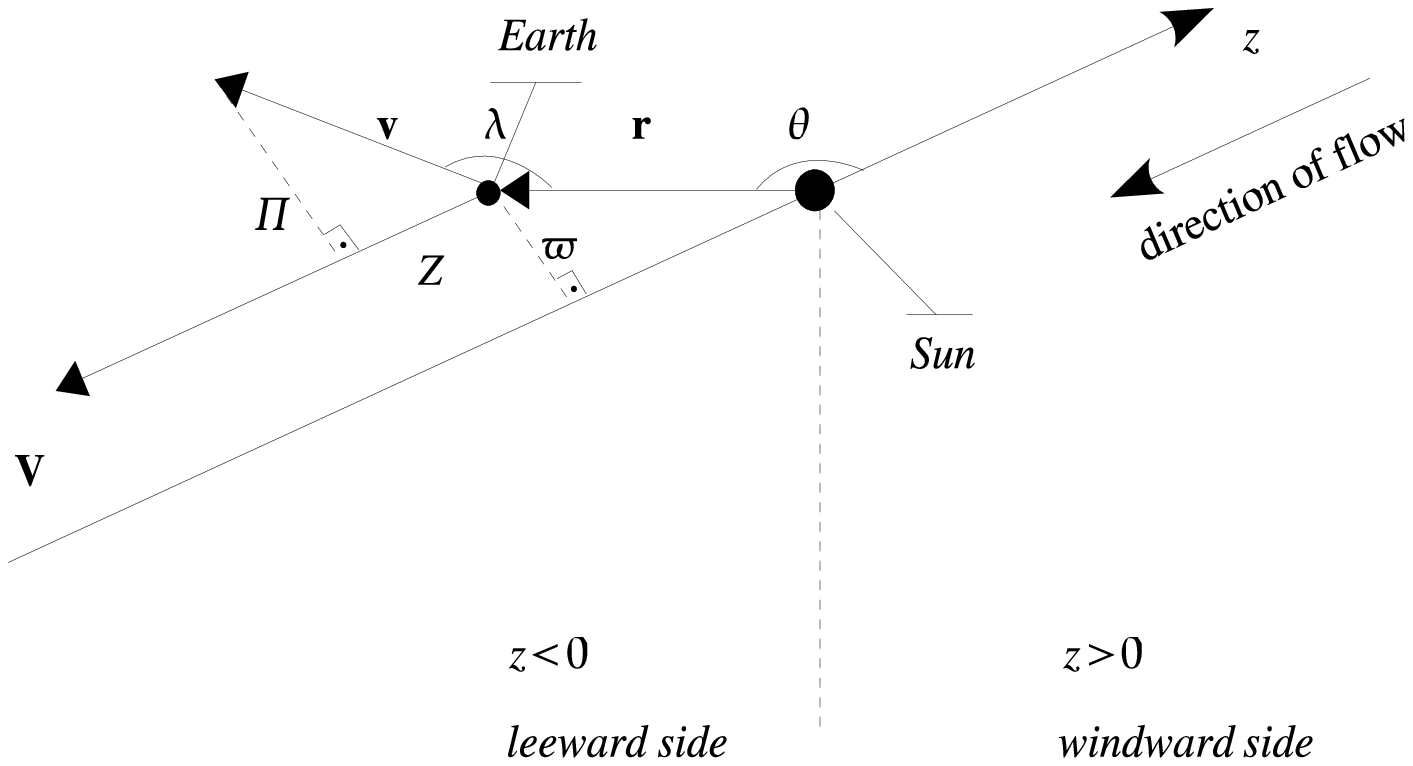}
\caption {Relating Danby and Camm's \cite{DC} and Griest's \cite{G} parameters to our parameters ${\bf r}$, ${\bf v}$, and ${\bf V}$.}
\label{DCpar}
\end{figure}

In the work of Danby and Camm \cite{DC}, the velocity distribution function is determined at all points around the Sun. Cylindrical polar coordinates $(\varpi,\varphi, z)$ are used with the origin at the center of attraction, and the $z$ axis directed towards the streaming cloud of particles (see Fig.~\ref{DCpar}). The coordinate $\varpi$ is the distance of the Earth from the flow axis, and $\varphi$ is the azimuthal angle. The corresponding velocity components are $ \Pi,\Phi, Z $. Danby and Camm's distribution function at any point reads, in their notation,
\begin{equation}
f ={{(2\pi)^{-3 / 2}}{\rho_{0}} {h^{3}}} \exp\!\left[{-\frac{1}{2}}h^2\left(c^2 + I_1 + 2c \frac{I_1 Z -\beta I_1^{1/2} \mu r^{-1} \cos\theta + \beta I_1^{1/2} Z^{2} \cos\theta + \beta I_1^{1/2} Z \Pi \sin\theta}{I_{1} + \mu r^{-1} + \beta I_1^{1/2} Z \cos\theta + \beta I_1^{1/2} \Pi \sin\theta}\right)\right] ,
\label{fDC}
\end{equation}
where
\begin{equation}
I_1=\Pi^2+\Phi^2+Z^2-\frac{2\mu}{r} .
\label{I_1}
\end{equation}
Since $\Pi$, $\Phi$, and $Z$ are the velocity components and $\mu = G M$, then $I_1$ is identical to our $v_{\infty}^2$ given by Eq.~(\ref {vinfty2}).

For the parameters $h$ and $\rho _0$, we have $h = 1/\sigma$ and $\rho _0 =\rho_\infty$.
The parameter $c$ in Eq.~(\ref{fDC}) can be obtained from the identity
\begin{equation}
\Pi^2 + \Phi^2 + (Z + c)^2 = \left| {\bf v} - {\bf V} \right|^2 ,
\end{equation}
which follows by comparing the exponents of the Maxwellian velocity distributions in Eqs.~(\ref{fDC}) and (\ref{eq:f}).
We find $c = V$. 

For the other parameters in Eq.~(\ref{fDC}), we use Fig.~\ref{DCpar} to define them in terms of our parameters and according to our notation.
In Fig.~\ref{DCpar}, $Z$ is the component of the particle's velocity ${\bf v}$ along the $z$ axis, $\Pi$ is the radial component of ${\bf v}$ (orthogonal to the $z$ axis), and $\theta$ is the angle between  the $+z$ axis and the position vector ${\bf r}$. (The angle $\lambda$ is used by Griest \cite{G} and will be discussed later when comparing our calculation to his results.) Since the velocity of the Sun ${\bf v}_s$ points in the $+z$ direction, the angle $\theta$ is also the angle between ${\bf r}$ and ${\bf v}_s$. On the windward side of the axis of symmetry (their $z>0$), $\sin\theta = 0$ and $\cos\theta = +1$. On the leeward side (their $z<0$), $\sin\theta = 0$ and $\cos\theta = -1$. From the rotational symmetry of the problem, the velocity distribution is independent of the azimuthal angle $\varphi$.    

Using ${\bf V} = -{\bf v}_s = - {\bf\hat z}$, we have
\begin{equation}
\cos \theta = {\bf\hat r} \cdot {\bf\hat z} = -{\bf \hat r} \cdot {\bf \hat V}
\label{costheta}
\end{equation}
and
\begin{equation}
Z = {\bf v} \cdot {\bf\hat z} = - {\bf v} \cdot {\bf \hat V} .
\label{Z}
\end{equation}

Since $\Pi$ is the radial component of ${\bf v}$, we can write it as
\begin{equation}
\Pi = {\bf v} \cdot {\bf\hat e}_{\varpi},
\label{Pidef}
\end{equation}
where ${\bf\hat e}_{\varpi}$ is a radial unit vector orthogonal to the $z$ axis. An expression for ${\bf\hat e}_{\varpi}$ valid off the flow axis is
\begin{equation}
{\bf\hat e}_{\varpi} = \frac{ {\bf\hat r} - {\bf \hat r} \cdot {\bf \hat V} \, {\bf \hat V} } {
| {\bf \hat r} - {\bf\hat  r} \cdot {\bf \hat V} \, {\bf \hat V} | } =
\frac{ {\bf \hat r} + \cos\theta \, {\bf \hat V} }{ \sin\theta } . 
\label{unitvec}
\end{equation}
The second equality follows from Eq.~(\ref{costheta}) and the fact that $\sin\theta > 0$ for $0 < \theta < \pi$. On the axis, $\sin\theta=0$ and Eq.~(\ref{unitvec}) contains a division by zero. However, Danby and Camm's expression contains only the product ${\bf\hat e}_{\varpi} \sin\theta$, which is well-defined. Indeed, their expression contains the combination $Z \cos\theta + \Pi \sin \theta$, which with the help of Eqs.~(\ref{Pidef}) and (\ref{unitvec}) becomes
\begin{equation}
 Z \cos\theta + \Pi \sin\theta = - {\bf v} \cdot {\bf\hat V} \cos\theta +
\frac{ {\bf v} \cdot {\bf \hat r} + \cos\theta \, {\bf v} \cdot {\bf \hat V}}{\sin \theta} \, \sin\theta = {\bf v}
\cdot {\bf \hat r} .
\label{Zcostheta}
\end{equation}

According to our analysis and notation, Eq.~(\ref{fDC}) therefore takes the form
\begin{equation}
f = \frac{\rho_\infty}{(2 \pi \sigma^2)^{3/2}} \exp\!\left(- \frac{F_{DC}}{2 \sigma^2}\right) ,
\label{eq:DC}
\end{equation}
where
\begin{equation}
F_{DC} = V^2 + v_\infty^2 - 2 {\bf V} \cdot  \frac{v_\infty^2 {\bf v} - \beta v_\infty (GM/r) {\bf\hat r} + \beta v_\infty {\bf v} ({\bf v} \cdot {\bf \hat r})}{v_\infty^2 + (GM/r) + \beta v_\infty ({\bf v} \cdot {\bf \hat r})} .
\label{DCDC}
\end{equation}
We used $I_1=v_\infty^2$, and $\cos\theta$, $Z$, and $Z\cos\theta+\Pi\sin\theta$ given in Eqs.~(\ref{I_1}), (\ref{costheta}), (\ref{Z}), and (\ref{Zcostheta}).

The quantity $F_{DC}$ can be put into the form
\begin{equation}
F_{DC} = V^{2} + v_{\infty}^{2} - 2 {\bf V} \cdot {\bf v}_{\infty} = \left | {\bf v}_\infty - {\bf V} \right|^2
\end{equation}
with ${\bf v}_\infty$ given by Eq.~(\ref{v}) provided the parameter $\beta=-1$. 

Danby and Camm introduced $\beta$ in Eq.~(\ref{fDC}) to deal with an ambiguity in the derivation of their formula. This ambiguity lead Danby and Camm \cite{DC} to confine their attention to the axis of symmetry instead of making full use of their expression. They solved their ambiguity by putting $\cos\theta=1$ and choosing $\beta=+1$ on the leeward side of the $z$ axis and $\beta=-1$ on the windward side. This choice gives the correct solution on the $z$ axis, because on this axis $\sin\theta=0$ and only the product $\beta \cos\theta$ appears in Eq.~(\ref{fDC}). With $\beta=-1$, the windward side has $\beta \cos\theta = -1$ and the leeward side $\beta \cos\theta = +1$. These are the same values that Danby and Camm obtain on the axis.

We conclude that Danby and Camm's expression is correct provided $\beta=-1$ and care is taken in using Eq.~(\ref{Zcostheta}) for $Z \cos\theta + \Pi \sin\theta$. Once this is done, their expression gives the same results as our numerical and analytical methods shown in Figs.~\ref{fig7}, \ref{vxphi}, and~\ref{par}.

\subsection{Danby and Bray}

Danby and Bray's distribution function, Eqs. (2) and (3) of their paper  \cite{DB}, reads, in their notation,
\begin{equation}
f =(2\pi)^{-3 / 2}{\rho_{0}} {h^{3}} \exp\!\left(\frac{-h^2}{2}F_{DB}\right) ,
\label{fDB}
\end{equation}
where
\begin{equation}
F_{DB} = c^2 + J^2 + 2c{\frac{J^2 Z - J(\mu/r) \cos\theta + J Z (Z \cos\theta + \Pi \sin\theta)}{J^2 + (\mu/r) + J(Z \cos\theta + \Pi \sin\theta)}}.
\label{expDB}
\end{equation}
Danby and Bray's notation is the same as Danby and Camm's, except that $J={I_1}^{1/2}$ and that the direction of the $z$ axis in Fig.~\ref{DCpar} has been reversed. 

Danby and Bray's formula is incorrect, as we now show. In our notation we have [cfr.\ Eqs.~(\ref{costheta}),(\ref{Z}), and (\ref{Zcostheta})]
\begin{eqnarray}
&& \cos\theta = {\bf \hat r} \cdot {\bf \hat V} , \\
&& Z = {\bf v} \cdot {\bf \hat V} , \\
&& Z \cos\theta + \Pi \sin\theta = {\bf v} \cdot {\bf \hat r} .
\end{eqnarray}
Hence,
\begin{equation}
F_{DB} = V^2 + v_\infty^2 - 2{\bf V} \cdot {\frac{- v_\infty^2 {\bf v} + v_\infty (GM/r) {\bf \hat r} - v_\infty {\bf v} ( {\bf v} \cdot {\bf\hat r})}{v_\infty^2 + (GM/r) + v_\infty ( {\bf v} \cdot {\bf\hat r})}} .
\end{equation}
This formula is incorrect because the $v_\infty^2 {\bf v}$ term in the numerator and the $v_\infty ( {\bf v} \cdot {\bf \hat r})$ term in the denominator have the wrong sign. To fix  Danby and Bray's formula, one should change the sign of the $J^2 Z$ term in the numerator and of the $J(Z \cos\theta + \Pi \sin\theta)$ term in the denominator in Eq.~(\ref{expDB}).

\subsection{Griest}

Griest \cite{G} used spherical coordinates with the Sun at the origin and the positive $z$ axis on the line from the Sun to the Earth. He considered a cloud of DM particles that, far away from the Sun, had a uniform density and an isotropic Maxwellian distribution of velocities.
Griest's distribution function \cite{G} reads, in his notation,
\begin{equation}
f =(2\pi)^{-3 / 2}{\rho_{0}} {h^{3}} \exp\!\left(\frac{-h^2}{2}F_{G}\right) \, \theta(J^2) ,
\label{fG}
\end{equation}
where
\begin{equation}
F_{G} = v_s^2 + J^2 + 2v_s {\frac{J^2 Z + J({GM_{\odot}}/a_{\oplus}) \cos\theta - J v Z \cos\lambda}{J^2 + ({GM_{\odot}}/a_{\oplus}) - J v \cos\lambda}} .
\label{expG}
\end{equation}
A typing mistake in Griest's paper \cite{G} was corrected in writing the above expression for $F_{G}$, namely $\cos\lambda$ in the last term in the denominator was mistakenly written as $\cos\theta$. This typing mistake can be seen by comparing Eqs. (2) and (5) of Griest's paper \cite{G}. 

The step function $\theta(J^2)$ in Eq.~(\ref{fG}) incorporates the assumption that there are no particles in bound orbits. In our calculation, we always have $\theta(J^2)=1$.

In the expression of $F_{G}$, $v_s$ is the Sun's velocity with respect to the cloud of DM, and according to our notation $v_s = V$. The parameter $J^2$ is identical to $v_{\infty}^2$, which is given by Eq.~(\ref {vinfty2}). The parameters $\cos\theta$ and $Z$ are the same as in Danby and Camm's paper \cite{DC} and related to our parameters by Eqs.~(\ref{costheta}) and~(\ref{Z}).
Furthermore, the parameter $v$ is the arrival speed of the DM particles on Earth with respect to the Sun, $a_{\oplus}$ is the distance between the Earth and the Sun, $h$ is the inverse of the  velocity dispersion, and $\rho_{0}$ is the density of the particles at infinity. Converting to our notation, we have $M_{\odot}=M$, $a_{\oplus} = r$, $h = 1/\sigma$, and $\rho_{0} = \rho_\infty$. Finally, from Fig.~\ref{DCpar} we can write
\begin{equation}
\cos\lambda = {\bf \hat v} \cdot {\bf \hat r} .
\label{coslambda}
\end{equation}

According to our analysis and notation, Griest's formula Eq.~(\ref{fG}) takes the form
\begin{equation}
f = \frac{\rho_\infty}{(2 \pi \sigma^2)^{3/2}} \exp\!\left(- \frac{F_{G}}{2 \sigma^2}\right) ,
\label{fG2}
\end{equation}
where
\begin{equation}
F_{G} = V^2 + v_{\infty}^2 - 2 {\bf V} \cdot {\frac{v_{\infty}^2 {\bf v} + v_\infty({GM}/r) {\bf \hat r} - v_\infty {\bf v} ({\bf v} \cdot {\bf \hat r})}{v_{\infty}^2 + ({GM}/r) - v_\infty ({\bf v} \cdot {\bf \hat r})}} .
\label{expG2}
\end{equation}
In writing the last expression, we used $J=v_\infty$, and $\cos\theta$, $Z$, and $\cos\lambda$ given in Eqs.~(\ref{costheta}), (\ref{Z}), and (\ref{coslambda}).

Eq.~(\ref{expG2}) is equivalent to
\begin{equation}
F_{G} = V^{2} + v_{\infty}^{2} - 2 {\bf V} \cdot {\bf v}_{\infty} = {\left| {\bf v}_{\infty} - {\bf V} \right|}^{2} 
\end{equation}
provided ${\bf v}_{\infty}$ is given by Eq.~(\ref{v}). Thus Griest's distribution function is identical to our expression for the Maxwellian distribution function, Eq.~(\ref{eq:f}), and it gives the correct distribution as in Figs.~\ref{fig7}, \ref{vxphi}, and~\ref{par}. 

\subsection{Sikivie and Wick}

We also compared our results to the work of Sikivie and Wick \cite{SW}. Sikivie and Wick assumed that the velocity distribution in the galactic halo is isothermal. They gave a position-dependent phase-space distribution in the presence of the Sun which in their notation reads
\begin{equation}
f({\bf r},{\bf v}) = \frac{d_{\infty}}{(\sqrt{\pi}\sigma)^{3}} \exp\!\left[-{\frac{1}{\sigma_{SW}^2}} v_{G}^{2}({\bf r},{\bf v})\right] \, \Theta(v _{\infty}^{2}({\bf r},{\bf v})) .
\label{fSW}
\end{equation}
As in Griest's paper \cite{G}, the step function $\Theta(v _{\infty}^{2}({\bf r},{\bf v}))$ incorporates the assumption that there are no particles in bound orbits. In our calculation, we always have $\Theta(v _{\infty}^{2}({\bf r},{\bf v})) = 1$. $d_{\infty}=\rho_\infty$ is the density of the particles at infinity, and $\sigma_{SW} = \sqrt {2/3} \sigma$. The factor $v_G^2({\bf r},{\bf v})$ in the exponent of Eq.~(\ref{fSW}) is given by \cite{SW}
\begin{equation}
v_{G}^{2}({\bf r},{\bf v}) = \bigl({\bf v}_{\odot} + {\bf v}_{\infty}({\bf r},{\bf v})\bigr)^{2} ,
\end{equation}
where ${\bf v}_{\odot} = - {\bf V}$ is the Sun's velocity with respect to the mean velocity of the DM particles, and ${\bf v}_{\infty}({\bf r},{\bf v})$ is given by the following formula in Sikivie and Wick's paper:
\begin{equation}
{\bf v}_{\infty}({\bf r},{\bf v}) = \frac{1}{a^{2} v_{\infty}^{2} + l^{2}} \left\{{\bf v} \left[l^{2} - a v_{\infty}^{2} r - a v_{\infty}({\bf r} \cdot {\bf v}) \right] + {\bf r} a v_{\infty}^{2} \left[\frac{1}{r}({\bf r} \cdot {\bf v}) + \frac{v^{2}}{v_{\infty}} - \frac{a v_{\infty}}{r} \right] \right\} .
\label{expSWv}
\end{equation}
Here $a = {G M} / {v_{\infty}^{2}}$, $l^{2} = r^{2} v^{2} - ({\bf r} \cdot {\bf v})^{2}$, and $v_{\infty}$ is given in Eq.~(\ref{vinfty2}).

The distribution function, Eq.~(\ref{fSW}), matches our DM particle distribution. Indeed, Sikivie and Wick's expression for ${\bf v}_{\infty}$, Eq.~(\ref{expSWv}), is analytically identical to ours, Eq.~(\ref{v}), due to the cancellation of the common factor
\begin{equation}
 v^{2}-\frac{GM}{r}+vv_{\infty} {\bf \hat v} \cdot {\bf \hat r} ,
\label{commonfactor}
\end{equation}
between the numerator and the denominator of their expression.

\section{CONCLUSIONS}
\label{conclusions}

In this paper, we analyzed the distribution of a flow of unbound collisionless DM particles in the Solar System, both numerically and analytically. In particular, we focused on the velocity distribution at the location of the Earth. We used Liouville's theorem to relate the phase-space distribution at the Earth to the velocity distribution at infinity (Sec.~\ref{distribution function}).

In the numerical method (Sec.~\ref{THE ARRIVAL DISTRIBUTION: NUMERICAL METHOD}), we traced the trajectories of the DM particles backward in time after they are deflected by the Sun's gravitational field. This numerical method is independent of the special form of the gravitational field and of the velocity distribution at infinity. The calculation used Newtonian gravity for advancing the trajectories until the particles return to infinity. The accuracy of the calculation was maintained by an adjustable time-step and a predictor-corrector iteration for improving the accuracy of each step along the trajectory.

In the analytical method (Sec.~\ref{derivation of v}) we obtained a formula, Eq.~(\ref{v}), for the velocity of the DM particles at infinity. That formula is valid for motion in a Keplerian field, and is independent of the choice of velocity distribution at infinity. 

We applied both the numerical and the analytic method to a selection of Maxwellian velocity functions at infinity representing streams of DM and the standard halo (Sec.~\ref{Phase-space distribution}). Comparison of the numerical and analytical calculations gave the same results.

For these velocity functions, we produced a number of sky maps showing the arrival distribution of DM particles at the Earth -- actually the specific flux defined in Eq.~(\ref{eq:N}) -- at different times of the year (see Figs.~\ref{fig7} and ~\ref{vxphi}). The maps also show the location of the Sun in the sky as the Earth moves around the Sun. The arrival distribution of DM particles under standard dark halo assumptions is displayed in Fig.~\ref{par}, which also shows the positions and magnitudes of the $1986$ brightest stars in the sky. 

The effect of the gravitational field of the Sun on the distribution of DM particles is evident on these maps. For example, the gravitational deflection produces a ring in the arrival distribution when the Earth is directly on the other side of the Sun as seen from the stream (see Fig.~\ref{fig7}(b)). This ring is analogous to the Einstein ring in the gravitational lensing of light. 

Finally, comparing our results with previous analyses, we were able to resolve the issue of discrepant results in  Danby and Camm \cite{DC}, Danby and Bray \cite{DB}, Griest \cite{G}, and Sikivie and Wick \cite{SW}. Danby and Camm's distribution function is correct on the axis of symmetry of the flow, and their ambiguity in the choice of the parameter $\beta$ can be fixed by choosing $\beta=-1$. Danby and Bray's have an incorrect distribution with a couple of wrong signs. Griest's distribution is correct after fixing a typo evident from comparing various equations in his paper. Sikivie and Wick's distribution is also correct. We checked that Danby and Camm's, Griest's, and Sikivie and Wick's distributions match the results of our numerical and analytical expressions shown in Figs.~\ref{fig7}, \ref{vxphi}, and~\ref{par}. 

\acknowledgments

The research described in this paper has been supported by a graduate student scholarship from the Saudi Ministry of Education, and by the National Science Foundation through grant PHY-0456825 at the University of Utah.


\begin{thebibliography}{9}

\bibitem{DK} T. Damour and L.M. Krauss, Phys.\ Rev.\ Lett. {\bf 81} (1998) 5726; Phys.\ Rev.\ {\bf D 59} (1999) 063509.

\bibitem{Gol} A. Gould, Astrophys.\ J. {\bf 368} (1991) 610.

\bibitem{GA} A. Gould and S.M.K. Alam, Astrophys.\ J. {\bf 549} (2001) 72.

\bibitem{LE} J. Lundberg and J. Edsjo, Phys.\ Rev.\ {\bf D 69} (2004) 123505.

\bibitem{DC} J.M.A. Danby and G.L. Camm, MNRAS {\bf 117} (1957) 50.

\bibitem{DB} J.M.A. Danby and T.A. Bray, Astron.\ J. {\bf 72} (1967) 219.

\bibitem{G} K. Griest, Phys.\ Rev.\ {\bf D 37} (1988) 2703.

\bibitem{SW} P. Sikivie and S. Wick, Phys.\ Rev.\ {\bf D 66} (2002) 023504.

\bibitem{gravitybook} B. Schutz, {\it Gravity from the Ground Up} (Cambridge Univ.\ Press, 2004).

\bibitem{WJ} W. Dehnen and J. Binney, Mon.\ Not.\ R. Astron.\ Soc.\ {\bf 298} (1998) 387.

\end{thebibliography}
\end{document}